# Performance Assessment of Resonantly Driven Silicon Two-Qubit Quantum Gate

Tong Wu and Jing Guo

*Abstract*—Two-qubit quantum gates play an essential role in quantum computing, whose operation critically depends on the entanglement between two qubits. Resonantly driven controlled-NOT (CNOT) gates based on silicon double quantum dots (DQDs) are studied theoretically. The physical mechanisms for effective gate modulation of the exchange coupling between two qubits are elucidated. Scaling behaviors of the singlet-triplet energy split, gate-switching speed, and gate fidelity are investigated as a function of the quantum dot spacing and modulation gate voltage. It is shown that the entanglement strength and gate-switching speed exponentially depend on the quantum dot spacing. A small spacing of ~10nm can promise a CNOT gate delay of <1 ns and reliable gate switching in the presence of decoherence. The results show promising performance potential of the resonantly driven two-qubit quantum gates based on aggressively scaled silicon DQDs.

## I. Introduction

QUANTUM computing promises attractive potential to solve certain problems not accessible to classical computers [1]–[3]. A two-qubit controlled-NOT(CNOT) gate and single-qubit rotational gates can form a complete set of quantum gates for universal quantum computing. Compared to the single qubit gates, physical realization of the two-qubit controlled gate is more challenging due to the need to create and control sufficient strong entanglement [4]–[13]. Significant experimental advances have been made to achieve two-qubit controlled quantum gates based on silicon double quantum dots (DQDs) [4]–[7]. Silicon-based quantum computing has the advantages of harvesting the well-established silicon fabrication infrastructure. Compatibility with CMOS technologies could lead to excellent scalability, high integration density, and low fabrication cost. Despite of recent pioneering experimental demonstrations, theoretical studies have mostly focused on fundamental studies on the exchange interaction without gate modulation [14] and designs for improved fidelity [15]. The performance potential and scaling behaviors of the resonantly driven, two-qubit quantum gates based on silicon, however, remain unclear.

In this letter, the gate modulation mechanisms, scaling behaviors, and performance potential of the resonantly driven CNOT gates based on silicon DQDs are assessed by using numerical device simulations. Three dimensional (3D) Poisson and Schrödinger equations are solved to model gate modulation of the exchange coupling. The qubit entanglement strength, gate switching speed, and effect of decoherence are studied as a function of the modulation gate voltage and DQD spacing. It shows that by scaling down the spacing between the gates that define the DQDs to ~10 nm, a short delay of <1 ns and high fidelity of >90% can be achieved for the resonantly driven silicon CNOT quantum gates. The results indicate promising performance potential of resonantly driven two-qubit quantum gates on silicon.

## II. Approach

The schematic structure of the modeled device is shown in Fig. 1(a). DQDs are defined by two side gates on a silicon-on-insulator (SOI) structure, with the middle gate modulating the strength of entanglement between two qubits [4], [7]. To explore the performance potential near the scaling limit, the thickness and gate size parameters are taken from aggressively scaled SOI or germanium-on-insulator (GOI) technologies [16], [17]. The silicon film thickness is $t_{si} = 2$ nm, and it is confined along [100] direction. The top gate insulator has a thickness of $t_{ins} = 3$ nm and a relative dielectric constant of $\kappa = 25$. The substrate is assumed to be 10 nm-thick $SiO_2$, with a ground plane. The temperature is assumed to be $T = 20$ mK. At this low temperature, dopants are frozen out, and the silicon film is effectively intrinsic. Valley degeneracy can potentially be an issue for silicon-based quantum gates. It, however, can be lifted by the interface effects [14], [18]. Only one valley, therefore, is considered in the energy range of interest.

In the experimentally demonstrated, resonantly driven CNOT gates on silicon, the entanglement between two qubits results in a singlet-triplet energy split and spin-dependent energy shift. A resonant microwave signal is used to drive spin-dependent switching [4]. In order to compute the singlet-triplet energy split, a configuration interaction (CI) method is used. The basis set of the CI method is obtained from the products of the lowest quasiparticle wave functions [14], [19], which are calculated by numerically solving the 3D Poisson and Schrödinger equations using the finite element method in the absence of the Coulombic interaction. The many-body Hamiltonian is then expressed in a matrix form, in which the quasi-particle part of the Hamiltonian is diagonal, and the two-body Coulombic interaction term introduces non-diagonal entries in the Hamiltonian matrix [19]. The lowest $N = 8$ quasiparticle eigenstates are used, which results in $N^2 = 64$ wave function products, as the basis set of the CI method. The many-body eigenstates and eigenvectors are then computed from an eigenvalue problem to obtain the singlet-

The authors are with the Department of Electrical and Computer Engineering, University of Florida, Gainesville, FL, 32611-6130 USA (e-mail: guoj@ufl.edu).

triplet energy split [19]. By examining its symmetry feature, a wave function can be assigned to be either a singlet state or a triplet state. The symmetric singlet ground state has a lower energy than the anti-symmetric triplet ground state. It is numerically tested that the CI basis set is sufficiently large for accurate calculation of the singlet-triplet split.

In the modeled, resonantly driven CNOT gate, a non-uniform magnetic field is applied, similar to the experiment [4]. In the presence of the magnetic field gradient, degeneracy of the triplet state is lifted, and four energy levels are formed. Fig. 1(b) shows the energy levels, $\{|\uparrow\uparrow\rangle, |\widetilde{\uparrow\downarrow}\rangle, |\widetilde{\downarrow\uparrow}\rangle, |\downarrow\downarrow\rangle\}$, where $E_z$ is the energy difference between the $|\uparrow\uparrow\rangle$ and $|\downarrow\downarrow\rangle$ states, and $dE_z$ is the Zeeman energy split due to the magnetic field gradient [4], [15]. In the modeled device, a DC-pulse voltage on the middle gate is combined with an AC resonant microwave signal to achieve the CNOT gate switching [4], [19]. The many-body adiabatic Hamiltonian is diagonal, and the resonant driving field results in the off-diagonal matrix elements of the Hamiltonian matrix $\widetilde{H}$. The microwave frequency is in resonance with the energy difference between the $|\uparrow\uparrow\rangle$ and $|\widetilde{\uparrow\downarrow}\rangle$ levels, so that this transition is resonantly excited. Other transitions are off resonance. As a result, a truth table of $\{|\uparrow\uparrow\rangle \rightarrow |\widetilde{\uparrow\downarrow}\rangle, |\widetilde{\uparrow\downarrow}\rangle \rightarrow |\uparrow\uparrow\rangle, |\widetilde{\downarrow\uparrow}\rangle \rightarrow |\widetilde{\downarrow\uparrow}\rangle, |\downarrow\downarrow\rangle \rightarrow |\downarrow\downarrow\rangle\}$ is obtained, which fulfills the CNOT gate functionality.

To obtain the transient characteristics of the resonantly driven CNOT gate, the Lindblad Master equation is solved [9], [10]

$$\frac{d\rho(t)}{dt} = \frac{-i}{\hbar}[\widetilde{H}, \rho(t)] + \sum_{k=1}^{N} \Gamma_k \left( O_k \rho O_k^+ - \frac{1}{2}\{O_k^+ O_k, \rho(t)\} \right), \quad (1)$$

where $\rho(t)$ is the time-dependent density matrix. On the right-hand side of the equation, the first term describes the coherent evolution. The second term describes the effect of decoherence phenomenologically, where the (m, n) matrix element of $O_k$ is $O_k(m,n) = \delta_{k,m}\delta_{k,n}$ with $\delta$ being the Kronecker delta function, $\Gamma_k = \gamma^*$, and $\gamma^*$ is the decoherence rate. dephasing of each eigenstate is treated, and the decoherence-induced transitions between different quantum states are neglected. A nominal value of the decoherence rate of $\gamma^* \approx (7.6 \text{ ns})^{-1}$ is used. Because of the long spin relaxation time due to weak spin-orbit and spin-phonon couplings and weak hyperfine interaction, a temperature of $T=20$ mK is sufficient to achieve a spin coherence time longer than this value [4], [7]. The decoherence model used here is simple and phenomenological, and it is used to examine the impact of decoherence on gate switching fidelity. This model does not provide a microscopic understanding on the physical mechanisms of decoherence.

### III. Results

The quasi-particle subband profile is presented first. For the modeled device as shown in Fig. 1(a), the ultrathin SOI film results in a strong quantum confinement in the vertical direction, which leads to two-dimensional (2D) subbands in the horizontal $x$-$y$ plane. The calculated lowest 2D subband at a side gate voltage of $V_{sL} = V_{sR} = 100$ mV is shown in Fig. 1(c), in which each quantum dot is populated with one electron. In the subsequent calculations, the side gate voltage is fixed, and the middle modulation gate voltage and the spacing $L_S$ are varied, in order to examine various issues on gate modulation and device scaling.

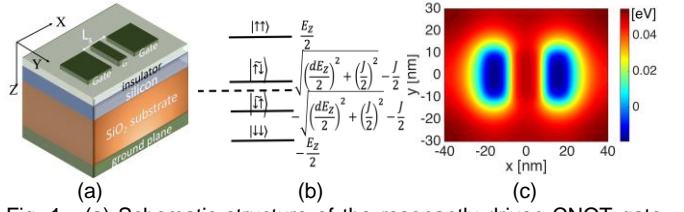

Fig. 1. (a) Schematic structure of the resonantly driven CNOT gate. Two quantum dots are defined electrostatically by the left and right side gates, whose entanglement can be modulated by the middle gate (G). The side gate spacing $L_S$ is denoted. $x = y = 0$ is defined at the center of the middle gate. The side gate size is 10 nm (along x) × 20 nm (along y). The middle gate size is $L_S$ minus 5 nm (along x) ×20 nm (along y), and it is symmetrically placed between two side gates. The silicon film is grounded. (b) The lowest eigenenergy levels in the presence of a non-uniform magnetic field. The dashed line denotes $E = 0$. (c) The lowest 2D subband energy profile $E_1(x,y)$ in eV. The spacing is $L_S = 20$ nm.

To examine the switching mechanisms of the quantum gate from the off-state to the on-state by applying a DC voltage on the middle modulation gate, the quasi-particle subband profiles along $x$ direction are plotted for the middle gate voltages of $V_{gm} = 20$ mV (off-state) and $V_{gm} = 70$ mV (on-state) in Fig. 2(a). It shows that an increase of the middle gate voltage by $\Delta V_{gm} = 50$ mV can result in a barrier height modulation of ~40 meV. The middle gate modulation on the middle barrier that separates two quantum dots is effective.

The strength of exchange coupling between two qubits is determined by the tunneling coupling and the Coulombic exchange integral term between the quantum dots,

$$I_X = \iint dr_1^3 dr_2^3 \psi_L^*(\vec{r}_1)\psi_R^*(\vec{r}_2)V_I(\vec{r}_1,\vec{r}_2)\psi_R(\vec{r}_1)\psi_L(\vec{r}_2), \quad (2)$$

where $V_I(\vec{r}_1, \vec{r}_2)$ is the Coulombic interaction kernel, and $\psi_L$ and $\psi_R$ are the quasi-particle Hartree-Fock wave functions of the left and right quantum dots, respectively. Both the Coulombic exchange integral and the tunneling coupling stem from the spatial overlap of the wave functions of the left and right quantum dots. Figs. 2(b) and (c) plot the products $\psi_L^*(\vec{r})\psi_R(\vec{r})$ in the off-state and on-state, respectively. The wave function overlap maximizes around $x = y = 0$, where the barrier modulation by the middle gate is most efficient. When the gate voltage increases by 50 mV from the off-state to the on-state, the wave function overlap increases exponentially, as shown in Fig. 2(c).

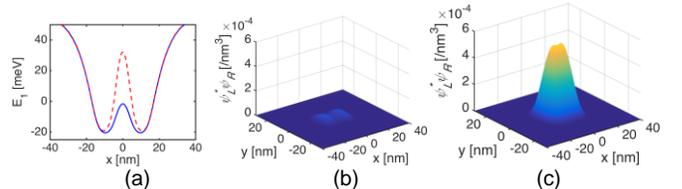

Fig. 2. (a) The subband profile along $x$ direction at y=0 and $z = t_{si}/2$ for a modulation gate voltage of 20 mV (dashed line) and 70 mV (solid line) at $L_S$=10 nm. The wave function overlaps in a horizontal cross section at a vertical position of $z = t_{si}/2$ for (b) the off-state $V_{gm} = 20$ mV and (c) the on-state $V_{gm} = 70$ mV.

The CNOT gate functionality relies on the entanglement between two qubits, which can be quantified by the singlet-triplet energy split, $J$. Fig. 3(a) shows the singlet-triplet energy split as a function of the voltage of the middle modulation gate $V_{gm}$ for different $L_S$ values. The energy split increases exponentially as the modulation gate voltage increases, because barrier lowering results in an exponentially larger overlap between the quasi-particle wave functions of the left and right quantum dots. In addition, Fig. 3(a) shows that as $L_S$ increases, the on-off ratio, $J(V_{gm} = 70 \text{ mV})/J(V_{gm} = 20 \text{ mV})$ increases. To further clarify the effect of $L_s$, the dependence of the singlet-triplet energy split on $L_S$ is plotted in Fig. 3(b) for the on-state and off-state. The singlet-triplet split $J$ exponentially increases as $L_S$ decreases [14]. As $L_S$ scales from 30 nm down to 10 nm, the value of $J$ increases by about 4 orders of magnitude. To achieve a sufficiently strong entanglement of $J$ >1 μeV at the on-state, a side-gate spacing of $L_S$ <20 nm is needed. On the other hand, a larger value of $L_S$ results in exponentially weaker entanglement between two quantum dots.

In an ideal switching event of a CNOT gate, the target bit is inverted when the control bit is 1, but it remains unchanged when the controlled bit is 0. To avoid erroneous switching, the Rabi frequency, $f_{\text{Rabi}}$, which is determined by the strength of the microwave signal, needs to be slow compared to $\sim J/h$, where $h$ is the Planck's constant. This requirement limits the switching speed of the resonantly driven CNOT quantum gates. We assume that the switching time is $\tau_{\text{switching}} = 10h/J$, which results in an erroneous inverting probability of <5% for the target bit when the controlled bit is set to 0. The switching delay time is plotted as a function of the side gate spacing $L_S$ in Fig. 4(a), for the on-state modulation gate voltage values of 60 mV and 70 mV. The results show that at $V_{gm} = 70$ mV, a switching delay of <1 ns can be achieved for a side gate spacing value of $L_S$ ≤12 nm. At $V_{gm} = 60$ mV, a switching delay of <1 ns can be achieved for a spacing value of $L_S$ ≤10 nm. As the quantum dot spacing increases, the gate becomes exponentially slower due to the exponential weaker entanglement between two qubits.

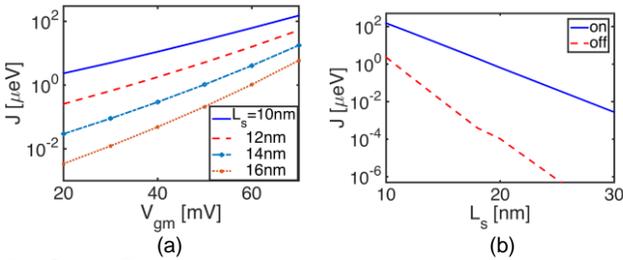

Fig. 3. (a) Singlet-triplet energy split vs. modulation gate voltage $V_m$ for the side gate distance of $L_S$ =10 nm (solid line), 12 nm (dashed line), 14 nm (dash-dot line), and 16 nm (dotted line), where $L_S$ is defined as shown in Fig. 1(a). (b) Singlet-triplet energy split vs. $L_S$ at $V_m$ = 70 mV (blue solid) and 20 mV (red dashed).

In the presence of decoherence, the fidelity of the quantum gate deteriorates significantly when the gate switching time becomes comparable to the decoherence time. Figs. 4(b) and (c) show the switching behaviors with and without decoherence for the resonantly driven CNOT gate with $L_S = 12$ nm and 20 nm, respectively. As discussed before, the effect of decoherence on the gate switching characteristics is modeled phenomenologically with a decoherence rate of $\gamma^* \approx (7.6 \text{ ns})^{-1}$. The results show that the gate fails to switch correctly at $L_S = 20$ nm, but it switches correctly at $L_S = 12$ nm. Creating strong entanglement by scaling down the side gate spacing, therefore, is advantageous for both faster switching and better immunity to decoherence.

Finally, we perform a quantum tomography analysis of the CNOT gate [20], and the fidelity is subsequently computed from the trace distance between the tomography matrix $\chi_{\text{ideal}}$ of an ideal CNOT gate and that of the modeled gate $\chi$, $F = 1 - \frac{1}{2}\sqrt{trace((\chi_{\text{ideal}} - \chi)^+(\chi_{\text{ideal}} - \chi))}$. A fidelity value of 92% and a short switching time of ~0.8 ns can be achieved for the side gate spacing value of $L_S$ =12 nm, because of the strong entanglement between two qubits. Due to the large value of $J$, the energy $dE_Z$ due to the magnetic field gradient is only comparable to $J$ for the modeled quantum gate with $L_S$=12 nm. While $dE_Z \gg J$ [4] is not a necessary condition for achieving reliable gate switching [21], [22], a large magnetic field gradient is still needed to produce $dE_Z \sim J$.

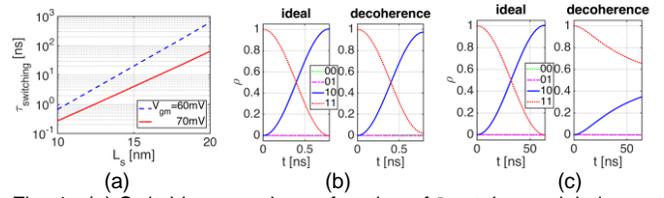

Fig. 4. (a) Switching speed as a function of $L_S$ at the modulation gate voltage values of $V_{gm}$ = 70 mV (the red solid line) and 60 mV (the blue dashed line). Transient switching characteristics at an applied gate voltage of $V_{gm}$ = 70 mV without decoherence and with decoherence for (b) $L_S$ =12 nm and (c) $L_S$ =20 nm. The modeled device structure is shown in Fig. 1(a). The state is initially prepared at |11⟩. The magnetic field is $B_L$=1.4 T at the left QD and $B_R$=1.0 T at the right QD. The effect of decoherence is modeled phenomenologically with a quantum state dephasing rate of $\gamma^* \approx (7.6 \text{ ns})^{-1}$.

## IV. Conclusions

Resonantly driven two-qubit quantum gates based on silicon DQDs are modeled by developing a 3D numerical device simulation capability. The results explain the physical mechanisms responsible for the efficient gate modulation on the qubit entanglement strength. Importance of scaling down the spacing between two quantum dots is highlighted for improving the gate switching speed and fidelity. For a spacing value of ~10 nm between the side gates, both fast switching of <1 ns and good fidelity can be achieved for the resonantly driven CNOT quantum gate. The results show the excellent performance potential of aggressively scaled, resonantly driven two-qubit quantum gates based on silicon.